# Prediction of a Two-dimensional Sulfur Nitride (S$_3$N$_2$) Solid for Nanoscale Optoelectronic Applications

Hang Xiao[1], Xiaoyang Shi[1], Feng Hao[1], Xiangbiao Liao[1], Yayun Zhang[1,3] and Xi Chen[1,2]

[1] *Columbia Nanomechanics Research Center, Department of Earth and Environmental Engineering, Columbia University, New York, NY 10027, USA*

[2] *SV Laboratory, School of Aerospace, Xi'an Jiaotong University, Xi'an 710049, China*

[3] *College of Power Engineering, Chongqing University, Chongqing 400030, China*

Two-dimensional materials have attracted tremendous attention for their fascinating electronic, optical, chemical and mechanical properties. However, the band gaps of most 2D materials reported are smaller than 2.0 eV, which greatly restricted their optoelectronic applications in blue and ultraviolet range of the spectrum. Here, we propose a new stable sulfur nitride (S$_3$N$_2$) 2D crystal that is a covalent network composed solely of S-N σ bonds. S$_3$N$_2$ crystal is dynamically stable as confirmed by the computed phonon spectrum and ab initio molecular dynamics simulations in the NPT ensemble. Hybrid density functional calculations show that 2D S$_3$N$_2$ crystal is a wide, direct band-gap (3.17 eV) semiconductor with good hole mobility. These fascinating electronic properties could pave the way for its optoelectronic applications such as blue or ultra-violet light-emitting diodes (LEDs) and photodetectors.

October 2004 marked the discovery of graphene [1], the first stable and truly 2D material. This epic discovery has opened up the possibility of isolating and studying the intriguing properties of a whole family of 2D materials including the 2D insulator boron nitride (BN) [2–4], graphane analogues of group IV elements, i.e. semimetallic silicene, germanene, and stanene [5–11], 2D transition-metal dichalcogenides [12–16], such as molybdenum disulfide [2,17,18] and tungsten disulfide [19], and very recently, 2D phosphorus, i.e. phosphorene [20], which extend the 2D material family into the group V. These 2D free-standing crystals exhibit unique and fascinating physical and chemical properties that differ from those of their 3D counterparts [21,22], opening up possibilities for numerous advanced applications. For example, MoS2, MoSe2, and WS2 are able to achieve 1 order of magnitude higher sunlight absorption than traditional photovoltaic materials such as GaAs and Si [23]. Two-dimensional materials offer novel opportunities for fundamental studies of unique physical and chemical phenomena in 2D systems [24,25].

In this work, we propose a new two-dimensional sulfur nitride (S$_3$N$_2$) solid (Fig. 1(a)) with space group Pmn21. The ground state structure of S$_3$N$_2$ was obtained using the evolutionary algorithm driven structural search code USPEX [26–28] combined with the ab initio code Quantum Espresso [29]. The S$_3$N$_2$ structure was further geometry optimized with density functional calculations with Perdew–Burke–Ernzerhof (PBE) [30] exchange-correlation functional using the Cambridge series of total-energy package (CASTEP) [31,32]. A plane-wave cutoff energy of 700 eV is used, and a 12 × 6 × 1 Monkhorst-Pack [33] *k*-point mesh was used. The convergence test of cutoff energy and *k*-point mesh has been conducted. Because the band gaps are dramatically underestimated by the GGA level DFT [34,35], band structures of S$_3$N$_2$ solid were calculated with HSE06 [36] hybrid functional, which has been demonstrated to be able to predict accurate band structures and density of states (DOS) [37]. All structure optimizations were conducted without imposing any symmetry constraints. The conjugate gradient method (CG) was used to optimize the atomic positions until the change in total energy was less than 5 ×10$^{-6}$ eV/atom, and the maximum displacement of atoms was less than 5 ×10$^{-5}$ Å.

The fully relaxed S$_3$N$_2$ crystal is depicted in Fig. 1(a). Despite its complex geometry configuration, the S$_3$N$_2$ crystal is a 2D covalent network composed solely of σ bonds (bonding is depicted by an isosurface of the electron density). The unit cell (inset to Fig. 1(a)) consists of ten atoms with lattice constants *a* = 4.20 Å, *b* = 8.92 Å. In the puckered structure, there are three types of S-N bond with bond lengths $d_1$ = 1.82 Å, $d_2$ = 1.73 Å, $d_3$ = 1.67 Å, and five types of bond angles $\theta_1$ = 117.2 °, $\theta_2$

= 118.9 ° and $\theta_3$ = 119.1 °, $\theta_4$ = 104.9 °, $\theta_5$ = 102.9 ° (depicted in Fig. 1(a)). Bond angles $\theta_1$, $\theta_2$ and $\theta_3$ are intermediate between the value corresponding to $sp^2$ bonding (120 °) and $sp^3$ bonding (109 °), while bond angles $\theta_3$ and $\theta_4$ are intermediate between the value appropriate for $sp^3$ bonding (109 °) and $p^3$ bonding (90 °). The Brillouin zone with the relevant high-symmetry k-points are depicted in the inset figure in Fig. 1(b).

By conducting phonon dispersion calculation of the free-standing $S_3N_2$ structure, we verify that all of phonon frequencies are real (Fig. 1(b)), confirming the dynamic stability of this structure. To further verify the stability of the structure at high temperatures, *ab initio* molecular dynamics (MD) simulations (shown in Fig. 2) at the PBE [30] /GTH-DZVP [38] level in the NPT ensemble of the CP2K [39] code. The simulations were run for 10 ps under ambient pressure at temperatures T= 800 K and 1000K, respectively. The stability of $S_3N_2$ structure is maintained at 800 K for 10 ps. However, the crystalline structure dissociates into multiple S-N chains and clusters at 1000 K. These MD calculations verify that the stability of $S_3N_2$ structure can be maintained well above the room temperature.

The band structure and density of states of the 2D $S_3N_2$ crystal are shown in Fig. 3. Calculations carried out by HSE06 hybrid functional show that the $S_3N_2$ structure is a semiconductor with a wide, direct band gap of 3.17 eV. This is a well-sought characteristic, since most 2D semiconductors reported thus far exhibit band gaps that are smaller than 2 eV. Both the valence band maximum (VBM) and the conduction band minimum (CBM) are composed of mainly the orbitals of sulfur atoms, as shown in Fig. 3. We also computed the effective mass of the electrons and holes (shown in Fig. 3) for the $S_3N_2$ structure at the $\Gamma$ point along the $\Gamma$-X and the $\Gamma$–Y directions. The quadratic band dispersion near the conduction band minimum is assumed when we estimate the effective mass of carriers. The effective electron masses were found to be $m_e^{\Gamma X} = 1.15\ m_o$ and $m_e^{\Gamma Y} = 1.26\ m_o$, where $m_o$ is the free-electron mass. The effective hole masses were obtained to be $m_h^{\Gamma X} = 0.75\ m_o$ and $m_h^{\Gamma Y} = 1.01\ m_o$. The effective mass of carriers along the $\Gamma$–X direction is lighter than that along the $\Gamma$-Y direction, showing an anisotropic transport property. Contrary to the common scenario where effective mass of hole is greater than electron, the hole effective mass in $S_3N_2$ crystal is lighter than its electron counterpart. The carrier mobility is expected to be good due to its small hole effective mass. As a 2D material with a wide, direct band gap, combined with a good carrier mobility, $S_3N_2$ crystal is ideal in optoelectronic applications such as blue or ultra-violet light-emitting diodes and semiconductor lasers. Furthermore, the band gap of $S_3N_2$ structure can be modulated by stacking or rolling up to form $S_3N_2$ nanotubes, opening up even more possibilities for its potential application.

In conclusion, we have presented a prediction of a new two-dimensional sulfur nitride ($S_3N_2$) material with distinguished structures and outstanding properties. Band structures calculated using HSE06 hybrid functional indicate that 2D $S_3N_2$ crystal is a semiconductor with wide, direct band gap of 3.17 eV. Our result not only marks the prediction of first sulfur nitride 2D crystal, but pave the way for exciting innovations in 2D electronics, optoelectronics, etc.

**FIGURES AND TABLES**

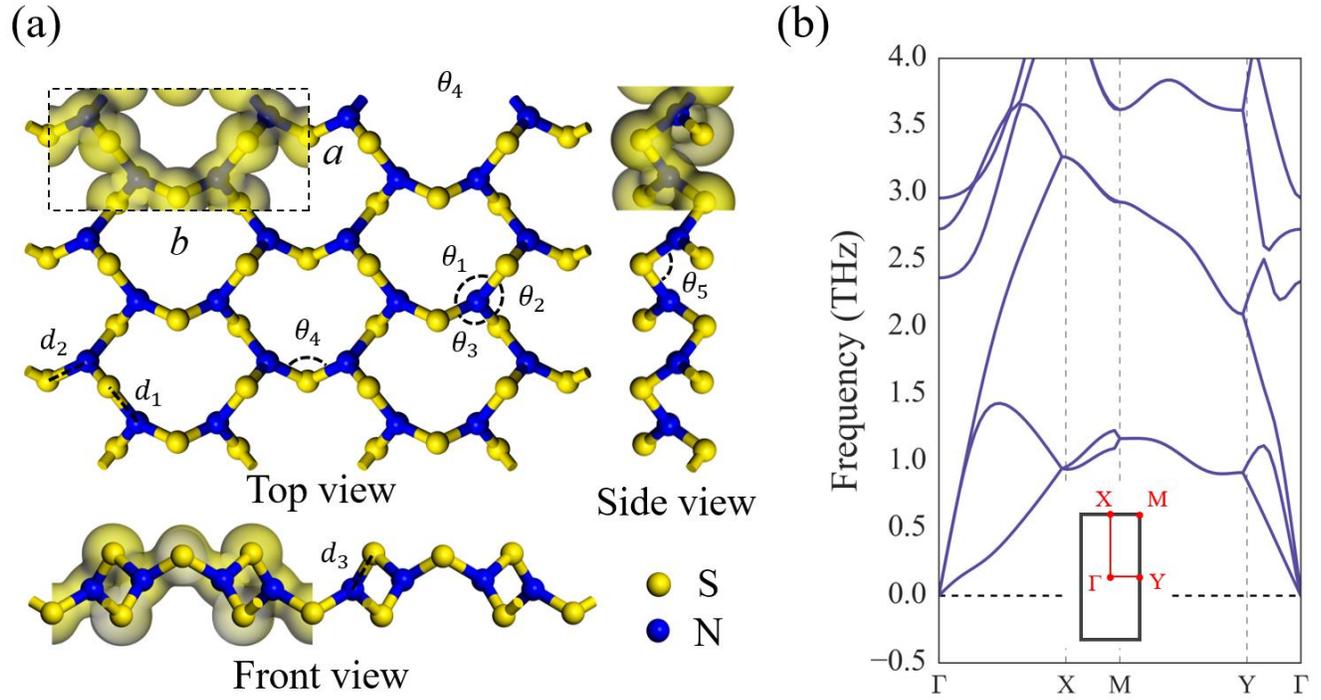

Fig. 1. (a) 2D crystalline structure (top view, front view and side view) of the sulfur nitride with lattice constants $a$ = 4.20 Å, $b$ = 8.92 Å, bond lengths $d_1$ = 1.82 Å, $d_2$ = 1.73 Å, $d_3$ = 1.67 Å, bond angles $\theta_1$ = 117.2 °, $\theta_2$ = 118.9 ° and $\theta_3$ = 119.1 °, $\theta_4$ = 104.9 °, $\theta_5$ = 102.9 °. Bonding is depicted by an isosurface of the electron density. (b) The phonon dispersion relation for the $S_3N_2$ solid. Structural stability is indicated by the absence of negative frequencies. The Brillouin zone, with the relevant high-symmetry k points indicated, is depicted in the inset figure.

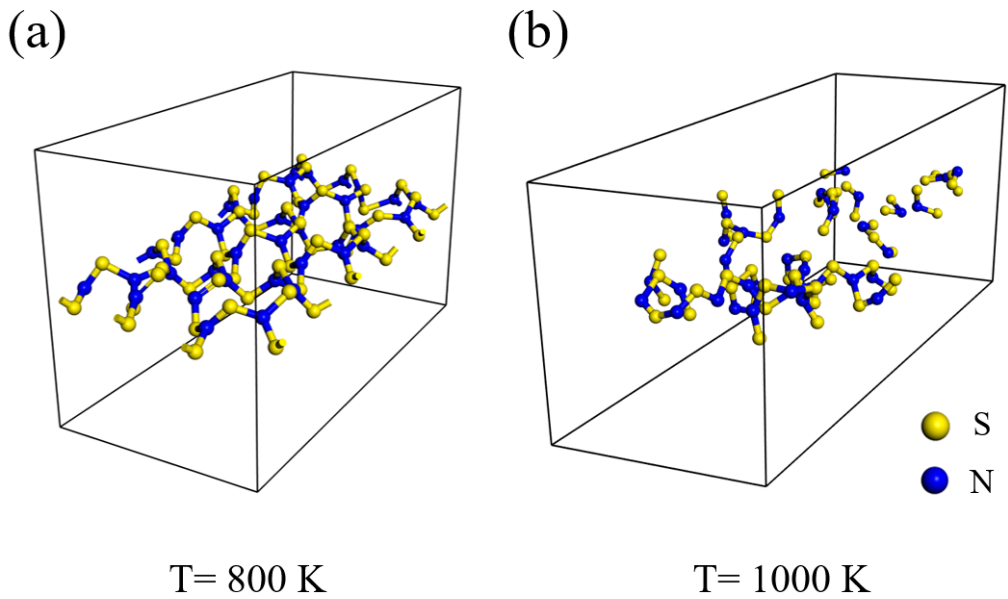

(a) T= 800 K  (b) T= 1000 K

Fig. 2. *Ab initio* MD snapshots of the $S_3N_2$ supercell structures at temperatures (a) T = 800 K (b) T = 1000 K under ambient pressure at 10 ps. The stability of $S_3N_2$ solid is maintained at 800 K, while at 1000 K the crystalline structure dissociates into multiple S-N chains and molecules. These MD simulations indicates the high degree of stability of $S_3N_2$ 2d crystal.

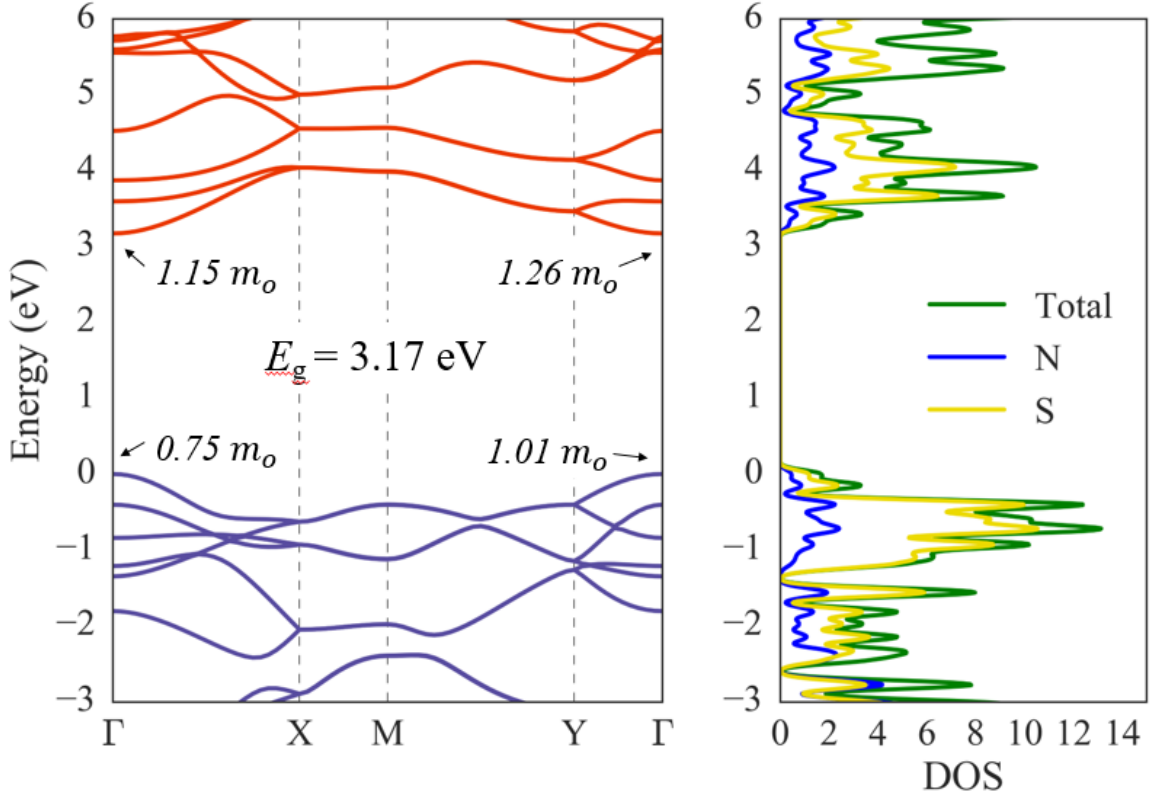

Fig. 3. Calculated band structure (left) and DOS (right) obtained with the HSE06 hybrid functional for the $S_3N_2$ solid. The effective mass of electrons and holes at the $\Gamma$ point along the $\Gamma$-X and the $\Gamma$-Y directions are indicated by black arrows, where $m_0$ is the true mass of an electron (9.11 $\times 10^{-31}$ kg). $S_3N_2$ crystal is a direct semiconductor with a band gap of 3.17 eV at the $\Gamma$ point. The hole effective mass along the $\Gamma$–X direction is only 0.75 $m_0$, indicating its good electric conductivity. Both the valence band maximum (VBM) and the conduction band minimum (CBM) are dominated by the orbitals of sulfur atoms, as shown in the DOS figure.

## ACKNOWLEDGEMENTS

The study is supported by DARPA (W91CRB-11-C-0112), AFOSR (FA9550-12-1-0159) and the National Natural Science Foundation of China (11172231).